\documentclass[sigconf]{acmart}
\usepackage{graphicx}
\usepackage{subcaption}
\graphicspath{ {./images/} }
\title{Empa: An AI-Powered Virtual Mentor for Developing Global Collaboration Skills in HPC Education}
\author{Ashish}
\affiliation{
  \institution{Purdue University}
  \city{West Lafayette}
  \state{IN}
  \country{USA}
}
\email{ashish@purdue.edu}

\author{Aparajita Jaiswal}
\affiliation{
  \institution{Purdue University}
  \city{West Lafayette}
  \state{IN}
  \country{USA}
}
\email{jaiswal2@purdue.edu}

\author{Sudip Vhaduri}
\affiliation{
  \institution{Purdue University}
  \city{West Lafayette}
  \state{IN}
  \country{USA}
}
\email{svhaduri@purdue.edu}

\author{Niveditha Nerella}
\affiliation{
  \institution{Purdue University}
  \city{West Lafayette}
  \state{IN}
  \country{USA}
}
\email{nnerell@purdue.edu}

\author{Shubham Jha}
\affiliation{
  \institution{Purdue University}
  \city{West Lafayette}
  \state{IN}
  \country{USA}
}
\email{jha70@purdue.edu}

\begin{document}

\begin{abstract}
High-performance computing (HPC) and parallel computing increasingly rely on global collaboration among diverse teams, yet traditional computing curricula inadequately prepare students for cross-cultural teamwork essential in modern computational research environments. This paper presents Empa, an AI-powered virtual mentor that integrates intercultural collaboration training into undergraduate computing education. Built using large language models and deployed through a progressive web application, Empa guides students through structured activities covering cultural dimensions, communication styles, and conflict resolution that are critical for effective multicultural teamwork. Our system addresses the growing need for culturally competent HPC professionals by helping computing students develop skills to collaborate effectively in international research teams, contribute to global computational projects, and navigate the cultural complexities inherent in distributed computing environments. Pilot preparation for deployment in computing courses demonstrates the feasibility of AI-mediated intercultural training and provides insights into scalable approaches for developing intercultural collaboration skills essential for HPC workforce development.
\end{abstract}


\maketitle
\section{Introduction}

Modern high-performance computing (HPC) and parallel computing environments are increasingly characterized by global collaboration, where researchers from diverse cultural backgrounds work together on complex computational problems ~\cite{Magana2022Teamwork}. From international climate modeling consortiums to distributed machine learning initiatives, successful HPC projects require teams that can navigate cultural differences in communication styles, work practices, and problem-solving approaches~\cite{olson2000distance, herbsleb2003introduction}. However, traditional computing curricula focus primarily on technical skills while inadequately preparing students for the cross-cultural collaboration and teamwork competencies essential in contemporary HPC careers~\cite{kumar2024computer,clear2025ai,garcia2025international,CS2023curriculum}.
The HPC workforce faces a documented challenge in developing globally competent professionals who can thrive in multicultural research environments ~\cite{barker2021multipronged}. Studies of international computational science projects reveal that cultural misunderstandings contribute to project delays, communication breakdowns, and reduced innovation when diverse teams cannot effectively collaborate~\cite{jirotka2013collaboration}. Furthermore, as HPC increasingly supports global challenges—from pandemic modeling to climate simulation—the ability to work across cultural boundaries becomes not just professionally advantageous but scientifically imperative~\cite{gill2024modern}.
Current approaches to developing cross-cultural collaboration competence in computing education remain limited and often disconnected from helping students understand the relevance of these skills to their future HPC careers. Traditional study abroad programs serve fewer students ~\cite{vande2023student} and rarely emphasize the connection between cultural competence and the effectiveness of collaboration and teamwork. Cultural awareness workshops, when offered, typically lack integration into computing curricula that would help students recognize their importance for careers in parallel computing, distributed systems, or computational science ~\cite{toti2025diversity}.
This paper presents Empa, an AI-powered virtual mentor designed to integrate intercultural collaboration training into undergraduate education, helping students develop skills to collaborate effectively with diverse others. The training modules are discipline-agnostic and applicable across all fields of study. In this work, we demonstrate Empa's deployment within computing curricula, emphasizing how intercultural collaboration skills are particularly critical for students who will enter careers requiring global teamwork, such as HPC and computational research environments. Our system leverages large language models to provide personalized, culturally sensitive guidance while students engage with collaborative scenarios that develop the multicultural teamwork skills increasingly essential across STEM disciplines. The system addresses three critical gaps in current undergraduate education: the lack of scalable intercultural collaboration training, the absence of curricular integration that helps students understand the relevance of cultural competence to their future careers, and the need for approaches that develop collaborative readiness alongside technical skills.
Our contributions include: (1) the design and implementation of an AI-mediated system for developing intercultural collaboration skills deployable across disciplines, (2) a framework for integrating general cultural competence training into technical curricula, (3) evidence of the feasibility and early effectiveness of AI mentors in developing global collaboration skills, and (4) insights into the challenges and opportunities for scaling cross-cultural competence training in undergraduate education, with particular focus on fields requiring global collaboration such as HPC.

\noindent{\bf Organization:} In this manuscript, we will first present the technical research origin (Section~\ref{techOri}), followed by the education adaptation and use case (Section~\ref{eduAdapt}). Next, in Section~\ref{archiOverview}, we will discuss the details of our system, including the LLM architecture, user interface, backend architecture, and module content and intended outcomes. Finally, we will present the curricular integration strategy (Section~\ref{currInte}), followed by insights from the instructor and students (Section~\ref{insights}), challenges and design tradeoffs (Section~\ref{challengesTradeoffs}), and future work and broader impacts (Section~\ref{FWBI}). 

\section{Technical Research Origins}\label{techOri}

The development of Empa builds upon established foundations in conversational AI for educational applications and intercultural competence theory. Large Language Models (LLMs) have demonstrated significant potential in educational contexts, with recent systematic reviews showing that AI tutoring systems can achieve learning gains exceeding traditional active learning approaches~\cite{kasneci2023chatgpt, hwang2023role}. However, the application of LLMs to cross-cultural competence training represents a relatively unexplored domain requiring careful integration of cultural frameworks with personalized educational guidance.
Our approach draws from two primary theoretical foundations. First, we leverage Hofstede's cultural dimensions theory~\cite{hofstede2010cultures}, particularly power distance, individualism versus collectivism, and time orientation, as fundamental frameworks for understanding cultural differences in collaboration and communication styles. Second, we incorporate Bennett's Developmental Model of Intercultural Sensitivity (DMIS)~\cite{bennett1993intercultural}, which provides a staged progression framework that guides the AI mentor's adaptive responses based on learner development in cross-cultural competence.
The technical architecture builds upon recent advances in prompt engineering and context-aware dialogue systems for educational applications. Empa implements personalized cultural scaffolding—a system design where the AI mentor maintains awareness of cultural frameworks to provide contextually appropriate guidance that helps students understand the relevance of intercultural skills to their future careers. This approach addresses limitations identified in prior work where educational AI systems struggled with providing personalized cultural competence training at scale~\cite{liu2023systematic}.
Our system design incorporates insights from Portable Intercultural Modules (PIMs) developed at Purdue University's Center for Intercultural Learning~\cite{jaiswal2023enhancing}. These discipline-agnostic modules provide the pedagogical foundation for structuring AI-mediated intercultural learning experiences. While the modules themselves are general, our deployment within computing curricula helps students understand how these universal collaboration skills are essential for success in global computing environments where cultural competence directly impacts project success and innovation outcomes.

\section{Educational Adaptation and Use Case}\label{eduAdapt}

Empa was designed to address a critical gap in computing education: the lack of structured, scalable opportunities for students to develop culturally responsive collaboration skills. While high-performance computing (HPC) and parallel computing increasingly rely on international teamwork, most curricula remain heavily focused on technical instruction, leaving little room for the development of interpersonal or intercultural competencies. Empa offers a lightweight, pedagogically aligned solution that can be integrated into computing courses without displacing core technical content.

The educational design of Empa emphasizes applied reflection through scenarios grounded in culturally diverse team interactions. Each of the six modules introduces students to core concepts—such as communication styles, power distance, individualism versus collectivism, and time orientation—through interactive media, guided reflection, and AI-generated feedback. Rather than focusing on technical workflow processes, Empa prompts students to examine how cultural values shape collaboration, conflict, and inclusion in global computing environments.

Use cases for Empa include embedding the modules as part of a unit on teamwork within computing courses, assigning them as pre-work for team-based projects, or incorporating them into ethics or professional development seminars. The system’s modular format enables instructors to assign one or more Empa activities at key points during the semester, for instance, before students begin working in groups or after a team milestone to help students reflect on their communication strategies, role expectations, and group dynamics.

Empa also supports differentiated learning by providing personalized, culturally contextualized feedback to each student. This ensures that learners receive guidance relevant to their own perspectives and experiences, promoting deeper engagement and self-awareness. Importantly, Empa’s framing is rooted in computing-relevant contexts, helping students see the relevance of intercultural collaboration skills to their future roles in global research or industry teams.

This educational adaptation positions Empa as a pedagogical bridge, supporting computing faculty in addressing interpersonal and intercultural competencies, while helping students cultivate the collaboration skills needed for success in increasingly multicultural HPC environments.

\section{System Architecture Overview}\label{archiOverview}

In this section, we will present the LLM architecture, followed by the user interface design and implementation. Next, we will present the backend architecture and its implementation. Finally, we will present our module content and intended outcomes. 

\subsection{Empa LLM Architecture}
The Empa LLM system is built upon the LLaMA 3 3B-Instruct architecture, which is a lightweight model that has been fine-tuned using parameter-efficient methods to optimize resource usage while ensuring high-quality feedback generation. Several studies have shown that fine-tuning lightweight models yield promising results with reduced compute resource requirements.

The training pipeline employs QLoRA, which is modular in nature and combines quantization with Low-Rank Adaptation (LoRA) in 4-bit precision. QLoRA introduces a number of innovations to save memory without sacrificing performance~\cite{dettmers2023qlora}. To enable efficient fine-tuning, we identify and target specific submodules that are suitable for LoRA injection by traversing the model architecture to collect the names of all unique submodules that are instances of bnb.nn.Linear4bit (excluding terminal layers such as \texttt{lm\_head}). This process ensures that LoRA adapters are only applied to internal linear layers relevant to model adaptation, thereby optimizing training efficiency and memory usage. This strategy aligns with the QLoRA paradigm, which combines 4-bit quantization with LoRA-based parameter-efficient fine-tuning. LoRA adapters are lightweight, trainable modules consisting of low-rank decomposition matrices inserted into the attention or feedforward layers of a frozen pre-trained model~\cite{hu2021lora}. Rather than updating all parameters—often numbering in the billions—LoRA enables fine-tuning by modifying only a small fraction of the model weights. By restricting injection to Linear4bit layers, we maintain quantization-aware efficiency while preserving the characteristics of the base model.

The training dataset comprises JSONL-formatted records, each containing a question, a synthetic student answer, a system prompt with tailored guidelines for feedback generation specific to the question, and the corresponding ground truth \texttt{empa\_feedback}. These system prompts are carefully designed to provide context-sensitive instructions that steer the model toward generating empathetic, constructive, and educational feedback. The inclusion of structured guidance within each prompt enables the model to condition its outputs effectively, improving consistency and relevance across diverse question-answer pairs. We implemented a custom masking strategy to ensure that the model learns to generate only the assistant's feedback (\texttt{empa\_feedback}), with loss computed exclusively on this portion. This is enforced using prompt formatting and label masking during dataset preprocessing, ensuring that student input and system instructions are present as context but excluded from loss computation. This setup prevents the model from being penalized for reproducing the input or instructional content, thereby aligning with the goal of high-quality, context-aware feedback generation.

Fine-tuning is conducted using Hugging Face's transformers and peft libraries. The PeftModelForCausalLM wrapper is used to inject LoRA adapters into specific transformer layers, and the adapters are trained while the base model remains frozen. Once training converges, the LoRA adapters are merged back into the base model using \texttt{merge\_and\_unload()}, producing the final Empa model suitable for deployment. The model is evaluated at fixed intervals per epoch using the Trainer API, with metrics tracked via Weights \& Biases. The entire system is configured for causal language modeling with a chat-format setup, allowing the model to produce structured, conversational feedback aligned with educational use cases.

This architecture enables Empa to efficiently learn relevant feedback patterns while maintaining a lightweight footprint for deployment. The fine-tuned model is deployed on Hugging Face in GGUF format, a lightweight and efficient binary format optimized for quantized models. Hosting on Hugging Face ensures easy access, reproducibility, and integration into downstream applications.

\begin{figure}[h]
\centering
\includegraphics[width=8.5cm]{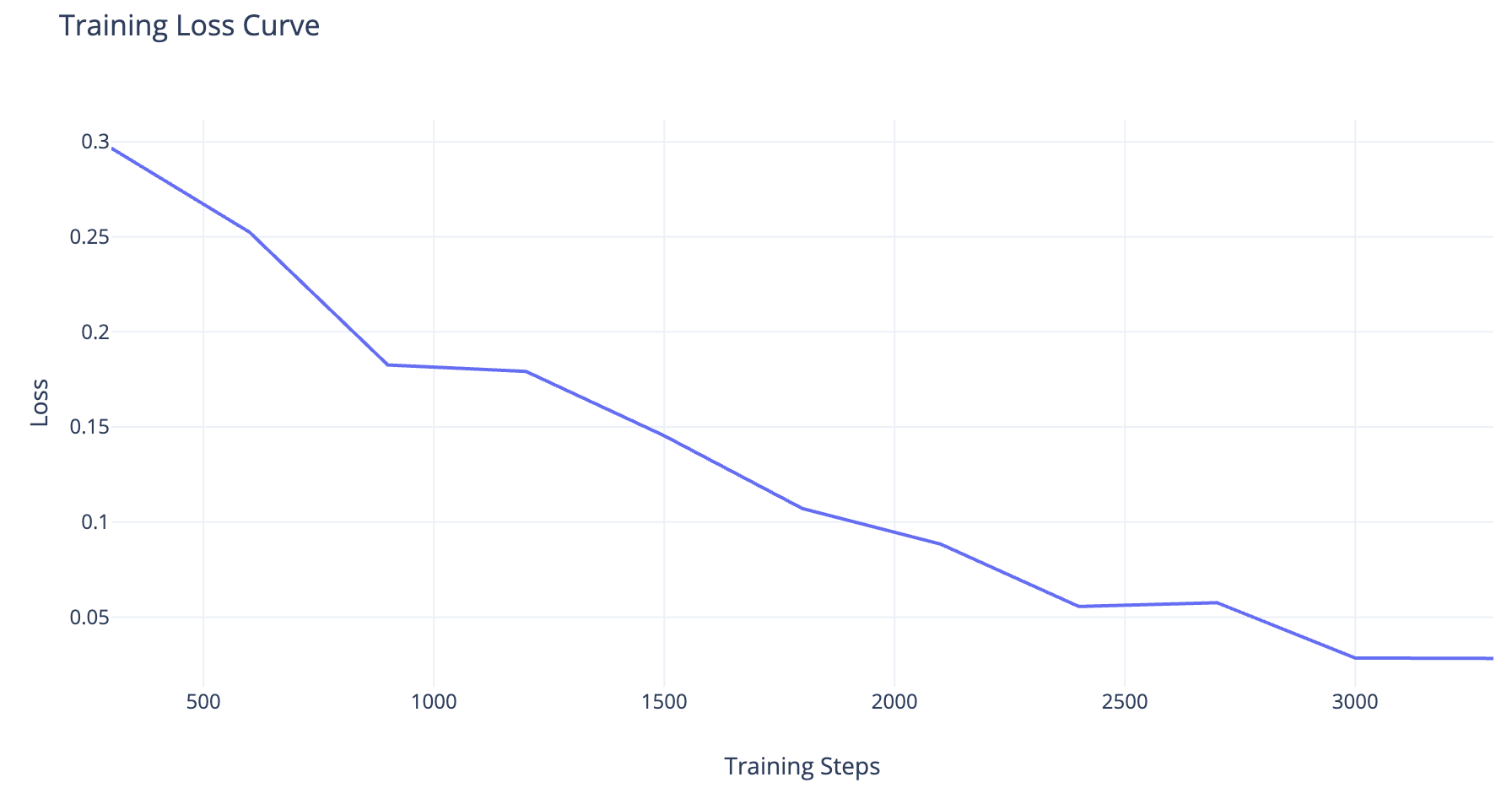}
\caption{Training loss curve showing convergence of the fine-tuned LLaMA 3 3B model during QLoRA fine-tuning.}
\label{fig:training-loss}
\end{figure}

\subsection{Empa User Interface}

The Empa user interface is designed to be modular, clean, and intuitive. It begins with a simple onboarding step where students enter their name and academic details, allowing the system to personalize the experience and refer to learners by name throughout the module.

The application follows a linear yet flexible navigation flow, beginning with onboarding and proceeding to six distinct sections: Exploring Interpersonal Collaboration, Meet Your Guide – Empa, Analyzing Team Interactions, Understanding Global Competence, Empathy as a Strategy, and Making Team Collaboration Work. Each section is structured to guide learners through increasingly deeper engagement with intercultural collaboration, using a combination of reflection prompts, visual media, AI-supported dialogue, and interactive quiz elements such as drag-and-drop cultural matching tasks.

The Empa platform is built as a modular, browser-based application using React and Tailwind CSS on the frontend, FastAPI for the backend, and Supabase as the cloud-based PostgreSQL database. This technology stack supports a responsive user experience with persistent session tracking and smooth data handling. The backend uses FastAPI to define lightweight RESTful endpoints that manage user registration, session state, and chatbot interactions. Supabase provides secure data persistence for both user profiles and chat history, enabling personalized continuity across sessions. Environment variables are used for secure configuration, and the modular API design ensures integration with the frontend components.

Key interactive features include:
\begin{itemize}
    \item \textbf{AI Chatbot:} A context-aware chatbot integrated into the right sidebar provides personalized feedback and reflection support, adapting to user inputs across modules.
    \item \textbf{Drag-and-Drop Quizzes:} Learners match characters to cultural categories, reinforcing concepts through applied, visual tasks.
    \item \textbf{Embedded Videos and Prompts:} Each module includes curated videos and structured reflection questions to support concept exploration.
    \item \textbf{Progressive Navigation:} Modules unlock sequentially based on completion, encouraging learners to move through the experience with increasing depth.
\end{itemize}

Together, these components enable an interactive and adaptive learning environment that supports students in building intercultural competence through guided reflection and applied practice.

\begin{figure}[ht]
  \centering
  \includegraphics[width=0.9\columnwidth]{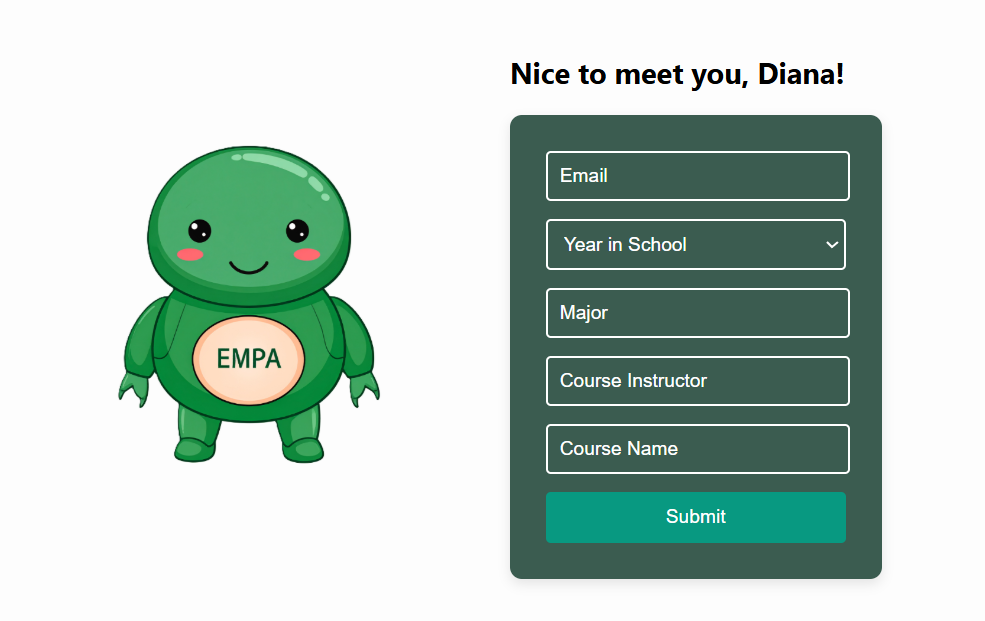}
  \caption{Onboarding screen where learners enter their academic details to personalize the Empa experience.}
  \label{fig:empa-login}
\end{figure}

\begin{figure}[ht]
  \centering
  \includegraphics[width=0.9\columnwidth]{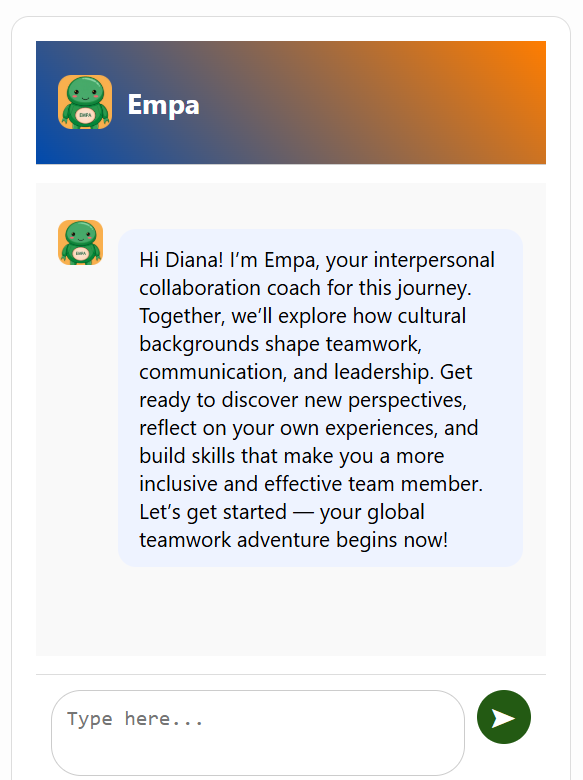}
  \caption{Empa chatbot greeting user with a personalized message and overview of the collaboration journey.}
  \label{fig:empa-chat}
\end{figure}

\begin{figure}[ht]
  \centering
  \includegraphics[width=0.9\columnwidth]{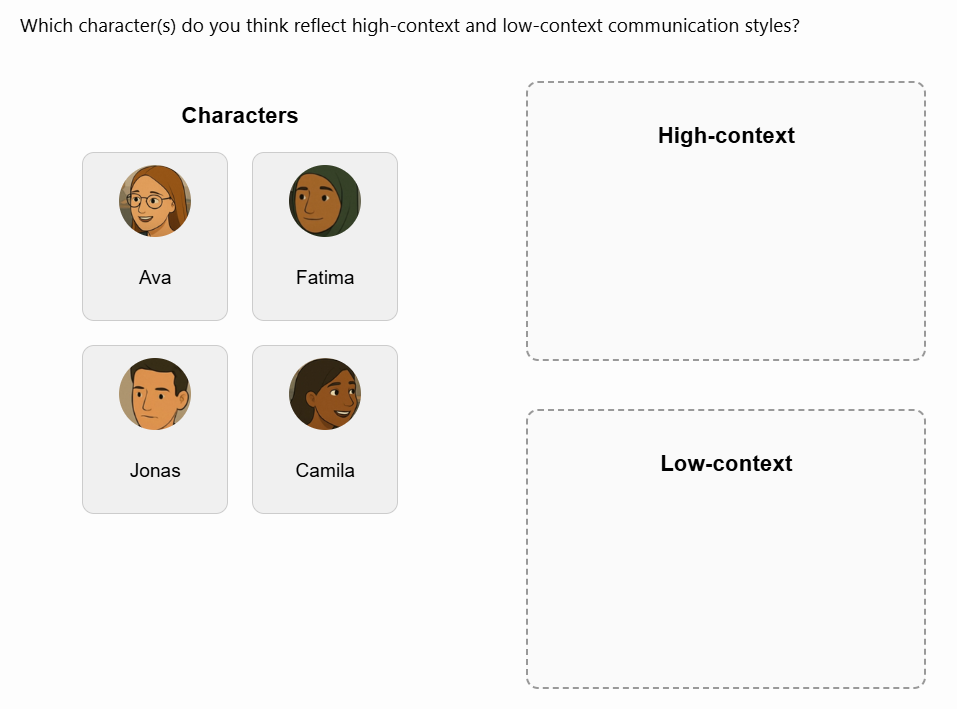}
  \caption{Drag-and-drop activity for classifying characters by communication style, reinforcing key cultural concepts.}
  \label{fig:dragdrop}
\end{figure}

\subsection{Backend Architecture and Implementation}

The Empa backend is designed as a RESTful API service that manages user sessions, persists conversational data, and orchestrates interactions between the frontend application and the large language model. Built using FastAPI, a modern Python web framework, the backend prioritizes performance, type safety, and developer experience while maintaining the scalability required for educational deployment.

\subsubsection{API Design and Endpoints}

The backend exposes three primary endpoints that support the complete user journey through the Empa learning modules:

\textbf{User Registration Endpoint (\texttt{POST /api/submit})}: This endpoint handles initial user onboarding by collecting demographic and academic information including name, email, year of study, gender, major, instructor, and course details. Upon successful registration, the system generates a unique user identifier and creates an initial personalized greeting message from Empa. This design decision ensures that each learner receives immediate, contextually relevant engagement that acknowledges their specific academic context.

\textbf{Conversational AI Endpoint (\texttt{POST /api/chatbot})}: The core interaction endpoint manages the stateful dialogue between learners and the Empa AI mentor. The endpoint maintains conversation context by retrieving the complete chat history for each user, formatting it according to the LLM's expected message structure, and appending new user messages before invoking the language model. This approach ensures temporal coherence in conversations while allowing the AI to reference earlier interactions to provide culturally contextualized guidance.

\textbf{Chat History Retrieval (\texttt{GET /api/chat-history/\{user\_id\}})}: This endpoint enables session persistence and conversation continuity by returning the complete interaction history for a given user. The chronologically ordered messages support both frontend state reconstruction and potential instructor review of student engagement patterns.

\subsubsection{Data Persistence Layer}

The backend utilizes PostgreSQL through Supabase, providing a robust and scalable cloud-based database solution. The schema design reflects the conversational nature of the application with two primary tables:

\begin{itemize}
    \item \textbf{Users table}: Stores learner profile information and demographic data essential for personalizing the intercultural learning experience
    \item \textbf{Chat\_history table}: Maintains a complete audit trail of all interactions, including timestamps, message content, and sender attribution (user or empa)
\end{itemize}

The database connection employs autocommit mode to ensure immediate persistence of all interactions, critical for maintaining conversation state across potential network interruptions common in educational environments. This design choice prioritizes data integrity and user experience continuity over transaction batching efficiency.

\subsubsection{LLM Integration Architecture}

The backend interfaces with the fine-tuned LLaMA model through a dedicated API service hosted on Purdue's GenAI infrastructure. The integration implements several key design patterns:

\textbf{Context Assembly}: Before each LLM invocation, the system reconstructs the complete conversational context by retrieving all previous messages and formatting them with appropriate role assignments (system, user, assistant). This ensures the AI mentor maintains awareness of the learner's progression through the modules and can provide developmentally appropriate feedback.

\textbf{System Prompt Injection}: Each conversation begins with a carefully crafted system prompt that establishes Empa's persona as a ``friendly, helpful, and knowledgeable'' mentor focused on interpersonal collaboration. This prompt engineering ensures consistent tone and pedagogical alignment across all interactions.

\textbf{Response Persistence}: Upon receiving the LLM's response, the system immediately persists both the user's query and Empa's reply to the database, maintaining conversation coherence even across session boundaries.

\subsubsection{Security and Scalability Considerations}

The backend implements several security best practices essential for educational deployment:

\begin{itemize}
    \item \textbf{CORS Configuration}: Explicitly defined origins ensure the API only accepts requests from authorized frontend deployments, preventing unauthorized access while supporting both local development and production environments
    \item \textbf{Environment Variable Management}: Sensitive configuration including database credentials and API keys are externalized through environment variables, following twelve-factor application principles
    \item \textbf{Input Validation}: Pydantic models enforce strict type checking and validation on all incoming data, with particular attention to email validation through the \texttt{EmailStr} type
\end{itemize}

\subsubsection{Error Handling and Resilience}

The implementation includes comprehensive error handling to ensure graceful degradation when external services are unavailable. The chatbot endpoint specifically handles GenAI API failures with appropriate HTTP status codes (502 for upstream errors), allowing the frontend to display meaningful error messages rather than cryptic failures. Database operations are wrapped in try-except blocks with detailed error reporting through traceback logging, facilitating debugging in production environments.

\subsubsection{Performance Optimization}

While the current implementation prioritizes simplicity and maintainability, several design decisions support future performance optimization:

\begin{itemize}
    \item \textbf{Stateless API Design}: Each request contains all necessary context, enabling horizontal scaling through load balancing
    \item \textbf{Prepared Statement Avoidance}: The explicit \texttt{prepare=False} parameter in database queries accommodates Supabase's connection pooling architecture
    \item \textbf{Minimal Processing Overhead}: The backend serves primarily as a thin orchestration layer, delegating complex computation to the LLM service and data persistence to the managed database
\end{itemize}

\subsection{Module Content and Intended Outcomes}

The Empa platform is structured into six sequential modules, each designed to deepen the learner's understanding of intercultural collaboration through a combination of media, reflection, and interaction.

\begin{itemize}
    \item \textbf{Exploring Interpersonal Collaboration:}  
    This opening module introduces the idea that different cultural backgrounds shape how individuals perceive the same situation. Learners reflect on a visual metaphor and short video illustrating multiple worldviews. 
    
    \textbf{\textit{Intended outcome:}} Learners develop awareness of cultural perception differences and begin building empathy by recognizing alternate interpretations of shared experiences.

    \item \textbf{Meet Your Guide – Empa:}  
    This module introduces learners to Empa, the AI chatbot. Through friendly dialogue, they explore Empa's role as a cultural mentor and reflective coach. 
    
    \textbf{\textit{Intended outcome:}} Learners become comfortable interacting with Empa as a personalized support tool and understand how it will assist in guiding cultural reflection.

    \item \textbf{Analyzing Team Interactions:}  
    Learners view a series of animated video scenarios representing the progression of a multicultural team over three weeks. 
    
    \textbf{\textit{Intended outcome:}} Learners begin to identify cultural tensions in team dynamics and reflect on how norms and communication styles influence collaboration.

    \item \textbf{Understanding Global Competence:}  
    This module introduces core cultural dimensions (e.g., power distance, communication style, individualism vs. collectivism, time orientation) through video explanations and interactive drag-and-drop quizzes that promote understanding through practical application and active engagement.
    
    \textbf{\textit{Intended outcome:}} Learners gain conceptual knowledge of key cultural frameworks and apply them through interactive sorting tasks to reinforce understanding.

    \item \textbf{Empathy as a Strategy:}  
    After exploring each cultural dimension, learners reflect on a new video where characters discuss their own cultural biases. Empa supports users in rethinking how they would respond in similar situations.  
    
    \textbf{\textit{Intended outcome:}} Learners strengthen their ability to reframe cultural conflict, practicing perspective-taking and empathy as tools for inclusive teamwork.

    \item \textbf{Making Team Collaboration Work:}  
    In the final module, learners reflect on a personal experience with team conflict and respond to a structured set of prompts. Empa offers coaching feedback based on their responses. 
    
    \textbf{\textit{Intended outcome:}} Learners synthesize their learning and apply cultural insight and empathy to a real-world scenario, building readiness for future collaboration.
\end{itemize}

\section{Curricular Integration Strategy}\label{currInte}

Empa was designed with seamless curricular integration in mind, recognizing that effective intercultural collaboration skills are critical yet underemphasized in undergraduate computing education ~\cite{bahrami2023information}. The system addresses the practical challenges instructors face when attempting to incorporate soft skills training into technically-focused curricula without sacrificing core content coverage or requiring extensive pedagogical restructuring.

The deployment architecture leverages a progressive web application that can be easily embedded into existing course structures through multiple integration pathways. Instructors can incorporate Empa modules as standalone assignments, pre-project preparation activities, or reflective exercises following team-based work. The modular design allows for flexible implementation—instructors may choose to deploy individual modules that align with specific course objectives or utilize the complete six-module sequence for comprehensive intercultural competence development.

Each Empa activity is structured as a low-lift, high-impact intervention that requires minimal instructor preparation while delivering measurable learning outcomes. The system provides context-specific prompts related to interpersonal communication, cross-cultural dynamics, power structures, and conflict resolution. While the core content remains discipline-agnostic, instructors can frame these activities within technical scenarios relevant to their courses, helping students understand the direct relevance of intercultural skills to their future careers in distributed computing, parallel programming, or computational research.

To support widespread adoption, Empa includes comprehensive instructor resources including plug-and-play assignment templates, learning objective alignments, and facilitation guides. These materials enable instructors to integrate intercultural learning without requiring specialized training in cultural competence pedagogy. The system's design philosophy prioritizes minimal disruption to existing syllabi while addressing core collaboration competencies that students will need in their professional careers.

The integration strategy emphasizes asynchronous participation and self-paced reflection, making Empa highly scalable across diverse institutional contexts and class sizes. Students can complete modules outside of class time, allowing instructors to maintain focus on technical content during class sessions while ensuring students develop essential collaboration skills. This approach accommodates both large lecture courses and smaller seminar-style classes without requiring different implementation strategies.

The system is currently being prepared for pilot deployment at Purdue University, where it will be tested in select computing courses to validate the integration approach and gather initial feedback from both instructors and students. The pilot implementation leverages existing Portable Intercultural Modules (PIMs) from the Center for Intercultural Learning, enhanced with Empa's AI-powered feedback system to create a more personalized and engaging learning experience. Following successful pilot validation, the system is designed for broader deployment across multiple courses and potentially other institutions, providing a scalable solution for intercultural competence training in technical education.

\section{Instructor and Student Insights}\label{insights}

Computing and HPC instructors face a well-documented challenge in developing students' interpersonal and intercultural collaboration skills.  Many instructors recognize the importance of mentoring students on communication styles, conflict resolution, and cultural awareness, yet struggle to provide personalized guidance at scale within large or technically demanding classes. The integration of such "soft skills" training often competes with time constraints and the need to cover extensive technical content ~\cite {chen2022producing}.

Empa addresses this instructional challenge by providing an AI-mediated layer of interpersonal coaching that can supplement faculty efforts without requiring additional class time or specialized instructor training in intercultural pedagogy. The system enables instructors to incorporate proven intercultural competence training into their courses through a scalable, automated approach that maintains focus on technical learning while developing essential collaboration skills.

From the student perspective, there is increasing recognition that technical expertise alone is insufficient for success in modern computing careers. Students entering fields like HPC, distributed systems, and computational research will inevitably work in multicultural teams where communication styles, cultural assumptions, and collaboration approaches directly impact project outcomes. However, traditional computing education provides few structured opportunities for students to develop awareness of how cultural differences affect teamwork dynamics or to practice navigating cross-cultural collaboration challenges.

Empa meets this educational need by providing students with personalized, reflective learning experiences focused on core cultural dimensions such as power distance, individualism versus collectivism, and communication styles. While the training content is discipline-agnostic, the deployment within computing curricula helps students understand the direct relevance of these skills to their future careers. The AI mentor provides culturally attuned feedback that supports skill development while fostering deeper self-awareness about personal collaboration patterns and cultural assumptions.

The system's strength lies in bridging the gap between general intercultural competence training and the specific needs of computing education. By providing scalable, personalized guidance that complements rather than competes with technical instruction, Empa offers a practical solution for developing globally competent computing professionals prepared for the collaborative realities of modern computational research and industry environments.

\section{Challenges and Design Tradeoffs}\label{challengesTradeoffs}

Developing Empa required navigating several design tradeoffs between technical constraints, pedagogical effectiveness, and scalability requirements. One major challenge was balancing cultural nuance with computational efficiency, especially as the model had to run effectively on lightweight architecture (LLaMA 3 3B-Instruct with QLoRA). To maintain responsiveness while generating context-rich, culturally sensitive feedback, we constrained the scope of feedback to 80-word windows and relied on careful prompt engineering for compression and focus. This limitation required extensive experimentation to ensure that brief responses could still convey meaningful cultural insights and actionable guidance.

Another significant tradeoff involved fidelity versus scalability: while deeper cultural training could potentially be achieved through extensive, long-form mentoring sessions, we opted for modular, activity-specific interventions that are easier to deploy and scale across multiple courses and institutions. This decision prioritized broad accessibility over depth of individual interactions, though the sequential module design still enables progressive skill development.

Aligning the AI mentor's tone and feedback style across diverse cultural contexts presented substantial complexity in system design. For example, providing appropriate feedback to students from cultures favoring direct communication versus those preferring indirect approaches required careful curation of training prompts and sophisticated masking strategies during fine-tuning to avoid cultural overgeneralization or bias. The challenge was ensuring that Empa could adapt its communication style while maintaining consistent educational objectives.

Ensuring the AI remained educational rather than performative or overly corrective also required extensive fine-tuning of loss masking and feedback formatting strategies. The system needed to provide constructive guidance that encouraged reflection and growth rather than simply correcting student responses. This balance was achieved through carefully designed training data that emphasized supportive, inquiry-based feedback over directive instruction.

Finally, the need to maintain system adaptability for future educational contexts and emerging pedagogical approaches required building an update-friendly architecture. The modular design and externalized prompt engineering enable ongoing refinement of the AI mentor's responses without requiring complete system retraining, supporting long-term sustainability and continuous improvement of the educational experience.

\section{Future Work and Broader Impacts}\label{FWBI}

Future work on Empa will focus on expanding its capabilities to support a wider range of collaborative learning scenarios, assessment features, and instructional integrations. One direction involves the development of adaptive team-based modules, where students can reflect not only as individuals but also receive aggregated feedback that surfaces group-level cultural dynamics. These enhancements aim to extend Empa's utility from self-reflection to peer and team development, supporting collaborative readiness at both the individual and group levels.

Another area for expansion is the inclusion of more nuanced cultural scenarios tailored to specific regions, institutional contexts, or academic levels. This will allow Empa to provide deeper cultural context while maintaining flexibility for use in different educational settings. Future development also includes planned integration with learning management systems (LMS) and data dashboards for instructors to enable more streamlined classroom deployment and facilitate evidence-based teaching practices.

We also envision developing a comprehensive formative assessment toolkit to help instructors evaluate growth in students' intercultural collaboration skills over time. This could include pre/post reflection prompts, behavioral indicators of culturally responsive teamwork, and rubrics for evaluating student responses. Such tools would provide instructors with concrete evidence of student development in intercultural competence, supporting both individual feedback and program-level assessment.

From a broader impact perspective, Empa represents a scalable, accessible intervention for embedding intercultural collaboration training into higher education—particularly in institutions that may lack resources for in-person intercultural programming or study abroad opportunities. By using AI to simulate personalized mentoring, Empa democratizes access to critical interpersonal skill development and promotes equity in workforce preparation across disciplines where global collaboration is essential.

The system's discipline-agnostic design positions it for adaptation across multiple STEM and non-STEM fields where multicultural teamwork is increasingly common. In the long term, Empa could serve as a model for culturally responsive AI mentors in diverse academic domains, from engineering and computer science to international business and public health. As professionals across fields engage in solving pressing global challenges—from climate modeling to public health informatics—the ability to work across cultural boundaries becomes a fundamental competency rather than an optional skill.

Empa contributes to a vision of higher education that values cultural humility, empathetic teamwork, and inclusive innovation as foundational competencies for students entering an increasingly interconnected global workforce. By providing scalable, evidence-based intercultural training, the system supports the development of graduates who are not only technically competent but also culturally responsive and prepared for the collaborative realities of modern professional environments.

\bibliographystyle{ACM-Reference-Format}
\bibliography{references}

\end{document}